\documentclass[number,sort&compress,5p,preprint]{elsarticle}

\usepackage{amsmath}
\usepackage{amssymb}

\begin{document}

\begin{frontmatter}

\title{Adaptive Monte Carlo applied to uncertainty estimation in five axis machine tool link errors identification with thermal disturbance.}

\author[poly,lurpa]{L.~Andolfatto\corref{loic}}
\ead{loic.andolfatto@lurpa.ens-cachan.fr}

\author[poly]{J.R.R.~Mayer}
\ead{rene.mayer@polymtl.ca}

\author[lurpa]{S.~Lavernhe}
\ead{sylvain.lavernhe@lurpa.ens-cachan.fr}

\cortext[loic]{Corresponding author. {Tel.:+33\,1\,47\,40\,27\,57}; {Fax:+33\,1\,47\,40\,22\,20}}

\address[poly]{Mechanical Engineering Department, \'Ecole Polytechnique de Montr\'eal, PO Box 6079, Station Centre-ville, Montr\'eal, Quebec, Canada H3C 3A7}

\address[lurpa]{Laboratoire Universitaire de Recherche en Production Automatis\'ee, \'Ecole Normale Sup\'erieure de Cachan, 61 Avenue du Pr\'esident Wilson, 94230 Cachan, France}

\journal{International Journal of Machine Tool and Manufacture}


\begin{abstract}
Knowledge of a machine tool axis to axis geometric location errors allows compensation and corrective actions to be taken to enhance its volumetric accuracy. Several procedures exist, involving either lengthy individual test for each geometric error or faster single tests to identify all errors at once.

This study focuses on the closed kinematic chain method which uses a single setup test to identify the eight link errors of a five axis machine tool. The identification is based on volumetric error measurements for different poses with a non-contact Cartesian measuring instrument called CapBall, developed in house.

In order to evaluate the uncertainty on each identified error, a multi-output Monte Carlo approach is implemented. Uncertainty sources in the measurement and identification chain -- such as sensors output, machine drift and frame transformation uncertainties -- can be included in the model and propagated to the identified errors. The estimated uncertainties are finally compared to experimental results to assess the method. It also reveals that the effect of the drift, a disturbance, must be simulated as a function of time in the Monte Carlo approach.

Results shows that the machine drift is an important uncertainty source for the machine tested.
\end{abstract}

\end{frontmatter}

\section{Introduction}

Five axis machine tools are well known in the industry for their high versatility and productivity by allowing the control of both position and orientation of the cutting tool relative to the workpiece. However, the higher number of axes results in more sources of errors, such as geometric link errors, which represent the relative location error between successive axes of a machine tool. Furthermore, the effects of those link errors are combined, resulting in potentially significant volumetric errors between the tool and the workpiece. Many studies have been carried out to estimate the link errors. Abbaszadeh-Mir et al. developed a linearised model relating the eight link errors of a five axis machine tool to the volumetric errors \cite{abba:2002} which can be solved in a least square sense to estimate the link errors. Lei and Hsu proposed a measuring instrument to evaluate volumetric errors \cite{lei:2002} with a method based on the closed chain principle \cite{benn:1991}. Bringmann and Knapp developed a method called "Chase-the-ball"  using a ball artefact mounted in the tool holder and linear probes to measure the Cartesian volumetric errors and identify link errors using a similar approach \cite{knap:2006}. Zargarbashi and Mayer presented a non-contact measuring instrument called CapBall to measure volumetric errors and identify the eight link errors \cite{zarg:2009}. The knowledge of those eight link errors allows to enhance the machine accuracy by a compensation for their effect on volumetric errors at the tool center point.

In \cite{knap:2002,knap:2009}, Bringmann and Knapp showed how the machine tool performance influence identification tests, considering the contribution of motion errors on uncertainty.

The purpose of this paper is to investigate the uncertainty contributions on the identified errors when using a Cartesian closed chain calibration approach and the CapBall sensing head. The identification procedure is briefly described and the result of identification are presented and evaluated. The second part presents the multi-output Monte Carlo technique and the sources of uncertainty considered.

The machine drift uncertainty is included in the model: a statistical model is compared to one that takes into account the cyclic character of the drift observed in the measurement chain.

Finally, the estimated uncertainties are compared to experimental results, and the uncertainty due to each source are separately evaluated to pinpoint the most penalising source of uncertainty.

\section{Relation between link errors and volumetric errors}

\subsection{Machine tool}
\begin{figure}
\center
\includegraphics[scale=1]{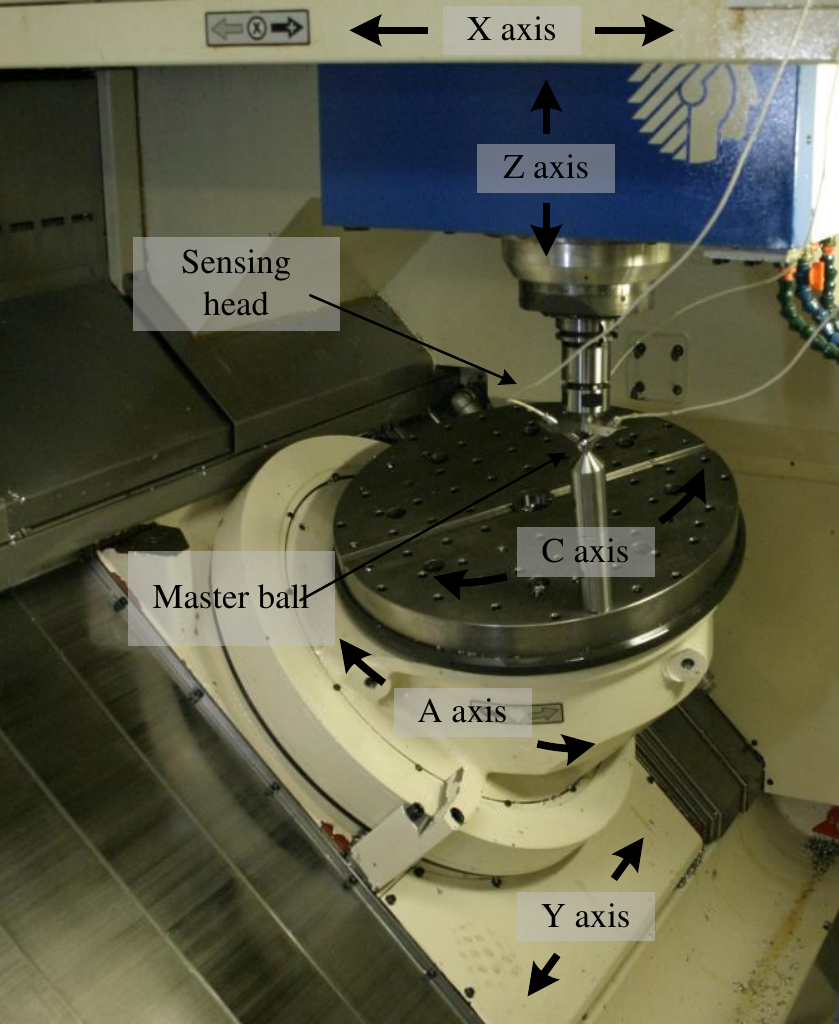}
\caption{CapBall system, composed of a sensing head attached in the spindle and a master ball mounted on the table.}
\label{fig:capball}
\end{figure}

The tests were performed on a Huron KX8-Five five axis machining center, depicted in Fig.~\ref{fig:capball}. 
This machine has a WCAYFXZT structure, with a 45-degree-tilted $A$ rotary axis. The strokes of the axis are $650\,mm$, $700\,mm$ and $450\,mm$ for $X$, $Y$ and $Z$ respectively. The numerical command controller is a Siemens Sinumerik 840D Powerline.

\subsection{Link errors model}

In \cite{abba:2002}, Abbaszadeh-Mir et al. presented a linear relation between the eight link errors of the machine and the six positioning errors of the setup -- gathered in an array $\boldsymbol{\delta p}$ -- on one hand and the three components of the Cartesian volumetric errors\footnote{The Cartesian volumetric errors are expressed as a vector containing the 3 components of the relative displacement of the tool center point to its nominal position.} at the tool center point expressed in the tool frame -- written in the vector $\boldsymbol{\delta \tau}$ -- on the other hand.

Following the methodology proposed in \cite{abba:2002}, the relevant error parameters are the following:

\begin{equation}
\begin{array}{ll}
\boldsymbol{\delta p} = & \left[ 
\delta \gamma_Y ~ \delta \alpha_Z ~ \delta \beta_Z ~ \delta \beta_A ~ \delta \gamma_A ~ \delta \alpha_C ~ \delta \beta_C ~ \delta y_C \right. \\
& \left. \delta x_T ~ \delta y_T ~ \delta z_T ~ \delta x_W ~ \delta y_W ~ \delta z_W \right]^T \end{array}
\label{eq:parameters}
\end{equation}

where $\delta \alpha_i$, $\delta \beta_i$ and $\delta \gamma_i$ denotes a small rotation of the joint $i$ around the X-, Y- and Z-axis respectively with respect to its nominal orientation (\emph{e.g.} $\delta \gamma_Y$ is the squareness between $X$ and $Y$ expressed as a small rotation of the $Y$ axis around the $Z$-axis); and $\delta y_C$ denotes a translation of joint $C$ in the direction $y$ with respect to its nominal position relative to joint $A$. Subscripts $W$ and $T$ stand for \emph{workpiece} and \emph{tool} respectively. Explanations about the components of the vector $\boldsymbol{\delta p}$ are given in Table~1.

The errors gathered in $\boldsymbol{\delta p}$ are modelled by small translational and rotational displacements. The Cartesian volumetric errors $\boldsymbol{\delta \tau}$ results from the sum of the effect of those small displacements, each propagated at the tool center point. Joint motion errors, \emph{e.g.} straightness, yaw, pitch, roll or positioning error of the axis are assumed negligible compared to the influence of link error in this model \cite{zarg:2009}.

The link errors are modelled with small displacement screw. Considering a pose $k$ of the machine, the small displacement screws can be transported at the tool tip involving linear transport equations. This leads to a linear operator $\boldsymbol{J}_k$ called \emph{Jacobian of the link and setup errors} and built as an algebraic expression of the sum of the effect of the previously introduced small displacements, leading to eq.\eqref{eq:jacobienne_defaut}.

\begin{equation}
\boldsymbol{\delta \tau} = \boldsymbol{J}_k \cdot \boldsymbol{\delta p}
\label{eq:jacobienne_defaut} 
\end{equation}

More details about the construction of the Jacobian are given in \cite{abba:2002} and \cite{zarg:2009}.

\section{Identification procedure}

\subsection{Measuring instrument}

In order to measure the volumetric errors at the tool center point $\boldsymbol{\delta \tau}$, a non-contact measuring instrument, called CapBall, had been developed in house \cite{zarg:2009}. It consists in a master ball mounted on the table and a sensing head fitted with three capacitive sensors, mounted in a tool holder in the spindle.

The three sensors have their axes nominally orthogonal and intersecting in one point called $P_t$. The machine is programmed to keep the tool center point $P_t$ coincident with the center of the master ball called $P_w$ (Fig.~\ref{fig:capball}).
Due to the geometric errors of the machine and setup, $P_t$ and $P_w$ are not exactly coincident, and the resulting volumetric error is measured by the sensing head.

The capacitive sensors are pre-calibrated for the spherical target using a set of three high precision linear stages. The response non-linearity is kept under $0.5\%$ as long as the eccentricity is kept within $\pm 300 \, \mu m$ and the distance between the sensor and the ball is in a $\pm 300 \, \mu m$ range centred on the position at which the sensor gives a 0 Volt signal.

\subsection{Identification}

Eq.\eqref{eq:jacobienne_defaut} provides three scalar equations for each pose of the machine. Those three scalar equations correspond to the measurement of the three components of the Cartesian volumetric error $\boldsymbol{\delta \tau}$ with the CapBall.

As the array $\boldsymbol{\delta p}$ has 14 components, it requires several poses to have enough scalar equations for its identification. With $n$ poses, eq.\eqref{eq:jacobienne_defaut} can be written $n$ times relating $n$ different values of the volumetric error vectors $\boldsymbol{\delta \tau}_1 \cdots \boldsymbol{\delta \tau}_n$ to $\boldsymbol{\delta p}$, as in eq.\eqref{eq:jacobienne_identification}.

\begin{equation}
\begin{bmatrix}\boldsymbol{\delta \tau}_1 \\ \vdots \\ \boldsymbol{\delta \tau}_n \end{bmatrix} = 
\begin{bmatrix}\boldsymbol{J}_1 \\ \vdots \\ \boldsymbol{J}_n \end{bmatrix} \cdot \boldsymbol{\delta p}
\label{eq:jacobienne_identification}
\end{equation}

The array $\boldsymbol{J}_{id}=\begin{bmatrix}\boldsymbol{J}_1 & \cdots & \boldsymbol{J}_n \end{bmatrix}^T$ is called the identification Jacobian. The column matrix $\boldsymbol{\delta \chi}=\begin{bmatrix}\boldsymbol{\delta \tau}_1 & \cdots & \boldsymbol{\delta \tau}_n \end{bmatrix}^T$ is called the error vector set. Finally, eq.\eqref{eq:jacobienne_identification} is written in a condensed manner as follows:

\begin{equation}
\boldsymbol{\delta \chi} = \boldsymbol{J}_{id} \cdot \boldsymbol{\delta p}
\label{eq:jacobienne_identification_short}
\end{equation}

The link and setup errors $\boldsymbol{\delta p}$ are identified using a Moore-Penrose pseudo inverse matrix:

\begin{equation}
\boldsymbol{\delta p}= {\boldsymbol{J}_{id}}^+  \cdot \boldsymbol{\delta \chi}
\label{eq:solution}
\end{equation}

A set of 807 poses has been defined as the identification trajectory. Those poses are chosen considering certain constraints:
\begin{itemize}
\item maximise the volume inscribed into the identification path to cover the largest part of the machine working envelop to identify more representative parameter estimates;
\item provide a relatively good condition number for the identification Jacobian to reach a better data noise immunity;
\item provide a large number of scalar equations to identify the parameters with a good outlier points robustness.
\end{itemize}

\begin{figure}
\center
\includegraphics[scale=1]{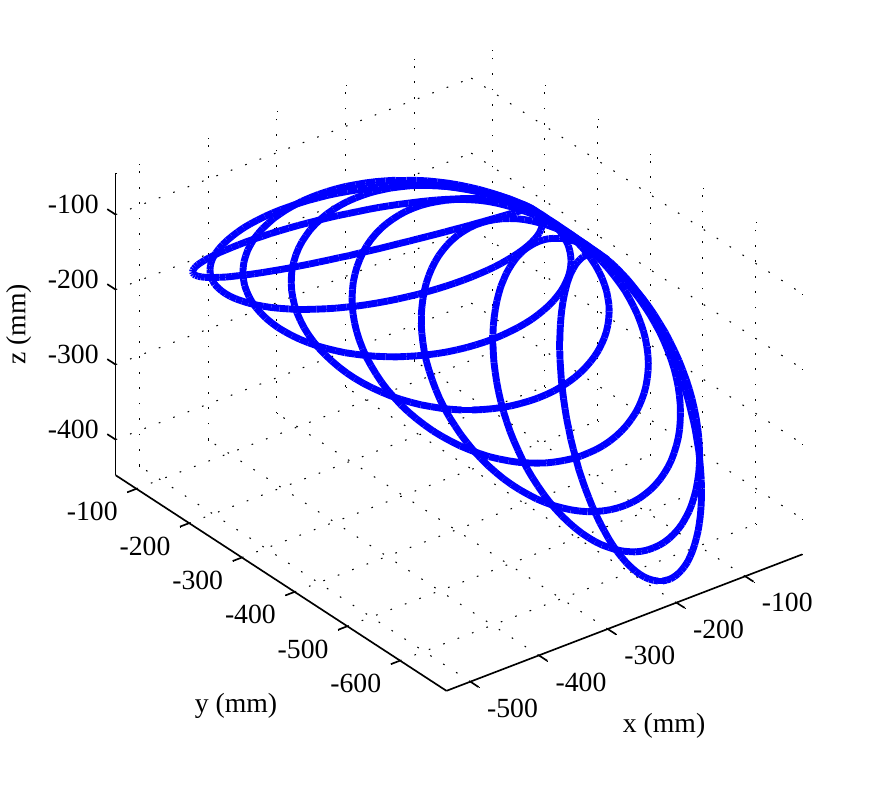}
\caption{Identification trajectory: successive positions of the points $P_t$ and $P_w$ in the machine coordinate system used to generate the identification Jacobian.}
\label{fig:trajectory}
\end{figure}

The trajectory shown in Fig.~\ref{fig:trajectory} describes the 807 successive positions of the points $P_t$ and $P_w$ in the machine coordinate system. The identification trajectory is performed with a clockwise A-axis and C-axis motion. The X-,Y- and Z-axis motions are calculated to keep the position of the tool tip relative to the workpiece nominally constant.

\subsection{Calibration of the sensors orientation}

During the measurements, the sensing head orientation is maintained constant relative to the machine coordinate system by programming the regulation of the spindle orientation on the machine. However, the sensing head measurement frame is not parallel to the machine tool frame.

The CapBall allows the measurement of the Cartesian volumetric errors in the sensing head frame. To be able to express $\boldsymbol{\delta \tau}$ in the machine tool frame, the orientation of the sensors axes must be calibrated.

The procedure is achieved by executing a calibration trajectory within a cube of $0.2 \,mm \times 0.2 \,mm \times 0.2 \,mm$ centred on the nominal position of the master ball, with a regular mesh of 125 points.
For each of the 125 points of the calibration trajectory, the vector $\boldsymbol{\delta \tau}_t$ defined from $P_w$ to $P_t$, expressed in the machine coordinate system, is programmed at a known value. At the same time, the sensor readings are recorded in $\boldsymbol{\delta \tau}_s$. The two following homogeneous coordinate matrices can be built to model this process:

\begin{equation}
\boldsymbol{\Delta}_t = \begin{bmatrix}
\boldsymbol{\delta \tau}_{t,1} & \cdots & \boldsymbol{\delta \tau}_{t,125}\\
1 & \cdots & 1
\end{bmatrix}_{4 \times 125} 
\end{equation}

\begin{equation}
\boldsymbol{\Delta}_s = \begin{bmatrix}
\boldsymbol{\delta \tau}_{s,1} & \cdots & \boldsymbol{\delta \tau}_{s,125}\\
1 & \cdots & 1
\end{bmatrix}_{4 \times 125}  
\end{equation}

Those two matrices can be theoretically related by a transform matrix $\mathcal{M}_{s \rightarrow t}$ to convert the sensors readings into master ball motion relative to the sensing head expressed in the machine frame:

\begin{equation}
\mathcal{M}_{s \rightarrow t} \cdot \boldsymbol{\Delta}_{s} = \boldsymbol{\Delta}_{t}
\end{equation}

Then, the best transform matrix from sensors frame to machine frame considering a least square criterion is:
\begin{equation}
\mathcal{M}_{s \rightarrow t} = \boldsymbol{\Delta}_{t} \cdot {\boldsymbol{\Delta}_{s}}^+
\label{eq:cal}
\end{equation}
where ${\boldsymbol{\Delta}_{s}}^+$ is the Moore-Penrose pseudo inverse of ${\boldsymbol{\Delta}_{s}}$.

With this method, none of the terms of $\mathcal{M}_{s \rightarrow t}$ are constrained. The sensors are ordered so that $(\boldsymbol{e}_1 , \boldsymbol{e}_2 , \boldsymbol{e}_3)$ is a right-handed reference frame. In an hypothetically perfect case, where the sensors directions would be exactly orthogonal, the gains perfectly known, the sensors response exactly linear and the machine movements perfect, then $\mathcal{M}_{s \rightarrow t}$ should be written as:

\begin{equation}
\mathcal{M}_{s \rightarrow t} = \begin{bmatrix}
\boldsymbol{e}_1 & \boldsymbol{e}_2 & \boldsymbol{e}_3 & \boldsymbol{d}\\
0 & 0 & 0 & 1
\end{bmatrix}
\end{equation}

\noindent
with the norm of each $\boldsymbol{e}_i$  equal to 1 and all the $\boldsymbol{e}_i$ forming together and orthogonal matrix. The experimental $\mathcal{M}_{s \rightarrow t}$ obtained with eq.\eqref{eq:cal} is quite similar to those theoretical values. The Details are given in Table~2. The last row of the experimental $\mathcal{M}_{s \rightarrow t}$ matrix is equal to its theoretical value, with an approximation of $10^{-15}$, supposed to be numerical errors only. 

Differences between theoretical and experimental norms of the $\boldsymbol{e}_i$ vectors come mainly from differences in the sensors gains. As shown in Table~2, it is less than 1\%. Considering the projections $\boldsymbol{e}_i \cdot \boldsymbol{e}_j$, the resulting values could be attributed to lack of orthogonality between the sensors axes and the machine's local out-of-squareness.

The vector $\boldsymbol{d}$ results from the sensors reading offset due to inaccurate axial position of the sensors in the sensing head. Each of its components were below $10\,\mu m$, which contributes to keeping the sensors in their linear domain.

The difference between the theoretical and the experimental $\mathcal{M}_{s \rightarrow t}$ is considered negligible. The experimental $\mathcal{M}_{s \rightarrow t}$ obtained during the calibration procedure is used for the identification of the link errors.

\subsection{Synchronism between poses and sensors values}

To measure the volumetric errors $\boldsymbol{\delta \tau}_k$ for the pose $k$, the trajectory is run with exact stop and dwell time at each pose with a tolerance of $1\,\mu m$. Compared to the method proposed by Zargarbashi and Mayer \cite{zarg:2009}, this requires a longer measurement time (about 10 minutes instead of 2), but considering the entire procedure, including preparation time, this is not a significant penalty.

The sensors signals are gathered continually trough a LABView application with a $1000\,Hz$ sampling rate. The volumetric errors for the programmed poses are extracted using a steady value algorithm developed for this purpose.

\begin{table}
\center
\caption{Result of identification -- link errors and setup errors.}
\label{tab:identification_result}
\vspace{4pt}
\small{
\begin{tabular}{p{6cm}c}
\emph{Link error}  &   \emph{Value}  \\
\hline
$\delta \gamma_y$, out-of-squareness between X and Y &    -8.8 $\mu m/m$  \\
$\delta \alpha_z$, out-of-squareness between Y and Z &    138.3 $\mu m/m$  \\
$\delta \beta_z$, out-of-squareness between X and Z &    -35.7 $\mu m/m$  \\
$\delta \beta_a$, tilt of A around $y$ &      -23.0 $\mu m/m$  \\ 
$\delta \gamma_a$, tilt of A around $z$ &   6.9 $\mu m/m$  \\
$\delta \alpha_c$, tilt of C around $x$ &    -34.4 $\mu m/m$  \\
$\delta \beta_c$, tilt of C around $y$ &    -9.9 $\mu m/m$\\
$\delta y_a$, offset of A relative to C in $y$ &   -2.9 $\mu m$ \\
\hline \hline 
\emph{Setup error}  &  \emph{Value}  \\
\hline
$\delta x_w$, ball (workpiece) setup error ($x$) 	&  	 1.5 $\mu m$ \\
$\delta y_w$, ball (workpiece) setup error ($y$)	&  -25,7 $\mu m$ \\
$\delta z_w$, ball (workpiece) setup error ($z$) 	&   18.8 $\mu m$ \\
$\delta x_t$, sensing head (tool) setup error ($x$) 		& 	-1.1 $\mu m$ \\
$\delta y_t$, sensing head (tool) setup error ($y$) 		&  -14.7 $\mu m$ \\
$\delta z_t$, sensing head (tool) setup error ($z$) 		&  -21.5 $\mu m$ \\
\hline
\end{tabular}}
\end{table}

\subsection{Identification result}

\begin{table}
\center
\caption{Comparison between theoretical and experimental values of $\mathcal{M}_{s \rightarrow t}$.}
\label{tab:matrice}
\vspace{4pt}
\small{
\begin{tabular}{p{3cm}ccc}
\emph{Norm} & $| \boldsymbol{e}_1 |$ & $| \boldsymbol{e}_2 |$ & $| \boldsymbol{e}_3 |$ \\ 
\hline 
Experimental & 0.998 & 0.996 & 0.996  \\ 
Theoretical & 1 & 1 & 1 \\
\hline 
\hline
\emph{Projection}  & $\boldsymbol{e}_1 \cdot \boldsymbol{e}_2$  & $\boldsymbol{e}_1 \cdot \boldsymbol{e}_3$ & $\boldsymbol{e}_2 \cdot \boldsymbol{e}_3$ \\
\hline 
Experimental & -0.030 & -0.029 & -0.021\\
Theoretical & 0 & 0 &  0 \\
\hline 
\end{tabular}}
\end{table}

\begin{figure*}[tbh]
\center
\includegraphics[scale=1]{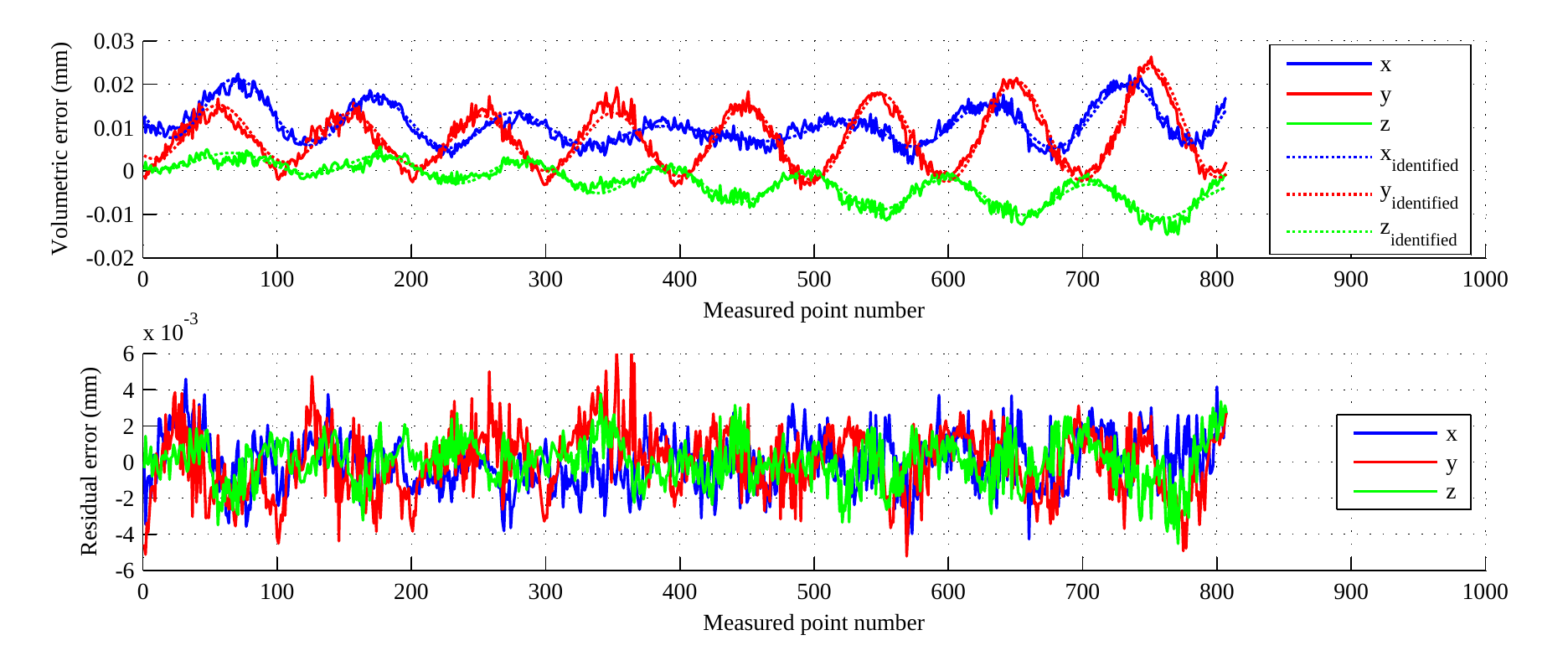}
\caption{Measured and predicted (according to identification) volumetric errors (top graph) and residuals only (bottom graph) along the identification trajectory.}
\label{fig:identification_result}
\end{figure*} 

\begin{figure}[tbh]
\center
\includegraphics[scale=1]{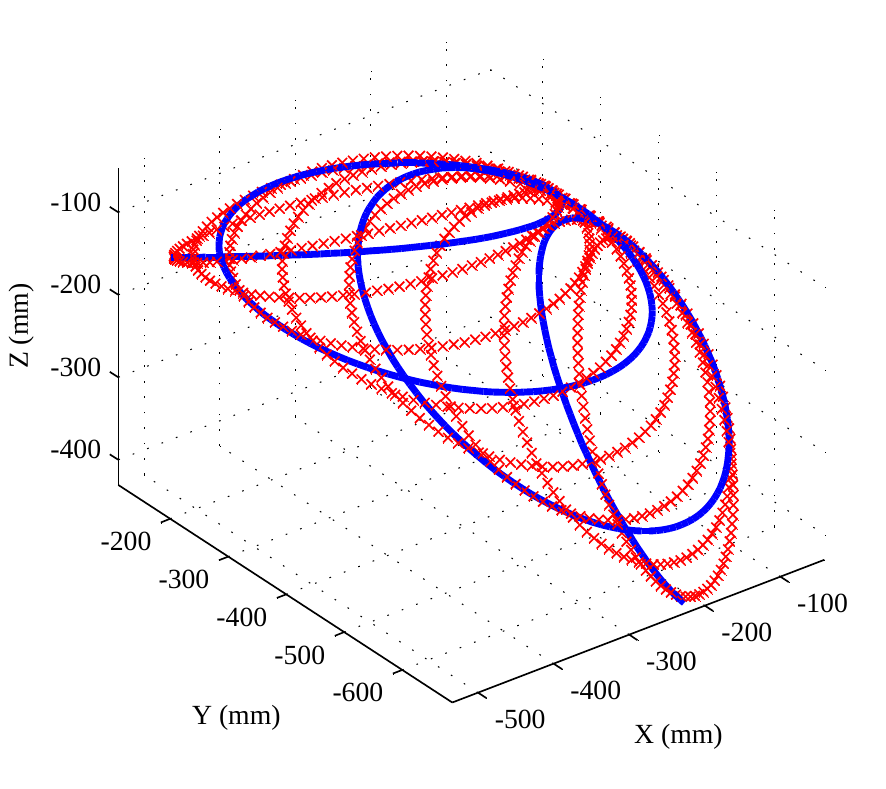}
\caption{Validation trajectory (blue) compared to identification trajectory (red).}
\label{fig:identification_trajectory}
\end{figure}

\begin{figure*}[bth]
\center
\includegraphics[scale=1]{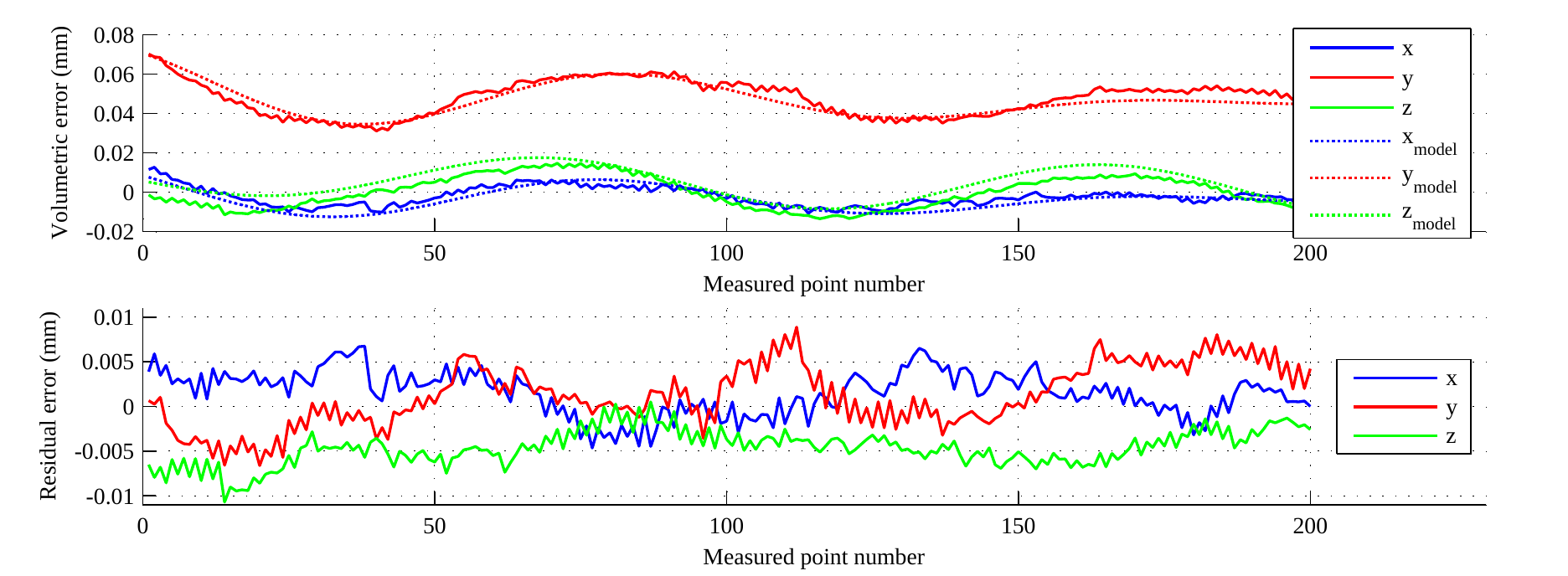}
\caption{Comparison of the experimental volumetric errors and model prediction for a validation trajectory.}
\label{fig:identification_validation}
\end{figure*}

The identification process has been performed using the identification trajectory illustrated in Fig.~\ref{fig:trajectory}. The identified error parameters are given in Table~1. 

Fig.~\ref{fig:identification_result} shows the measured volumetric errors and the prediction using the identified model (\emph{top}). The difference between measured errors and modelled errors is called \emph{residual errors} (\emph{bottom}). This portion of the errors unexplained by the model are under $10 \, \mu m$, with root mean square value of $1.5\, \mu m$, $1.8\, \mu m$ and $1.3\, \mu m$ in the $x$, $y$ and $z$ directions respectively.

\subsection{Validation of the identified errors}

The robustness of the identified model is evaluated by comparing its predictions against experimental measurements for a validation trajectory (Fig.~\ref{fig:identification_trajectory}) run with a counter-clockwise A-axis motion. The comparison between predictions and measurements is given in Fig.~\ref{fig:identification_validation}. The RMS values of the residual errors are then equal to $2.7\, \mu m$, $3.6\, \mu m$ and $2.5\, \mu m$ in $x$, $y$ and $z$ direction respectively. Those values are twice as large as those for the identification trajectory.

\section{Monte Carlo analysis}

\subsection{Structure of the simulation}

\begin{figure*}[tbh]
\center
\includegraphics[scale=1]{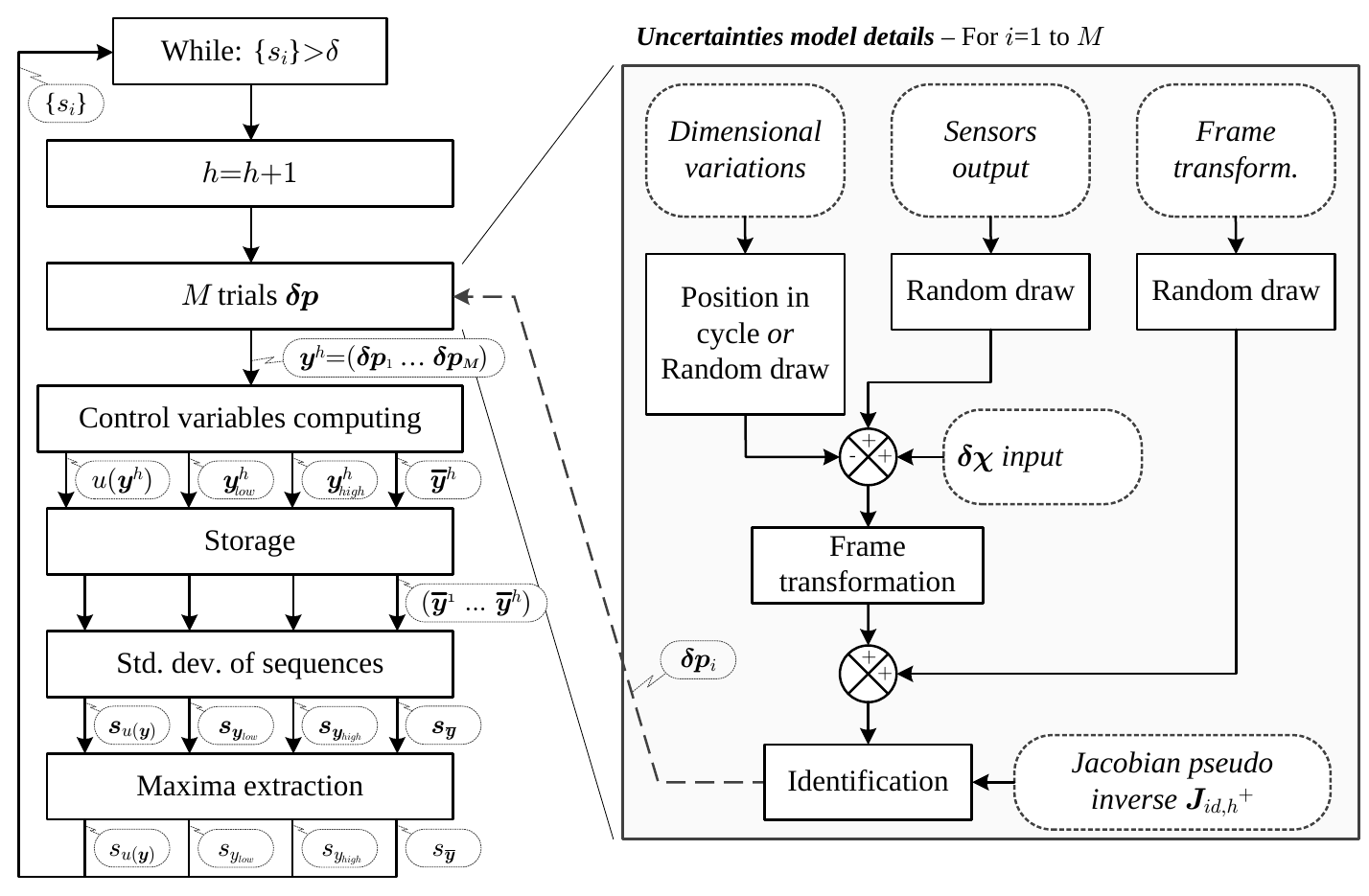}
\caption{Structure of the implemented Monte Carlo method as described in \cite{gum_mc}.}
\label{fig:structure_mc}
\end{figure*}

A Monte Carlo approach has been implemented to evaluate the uncertainty of the identified link errors. For such problems of least square identification, a Monte Carlo approach is not necessarily needed if uncertainties on the input can be modelled by a simple distribution \cite{kyri:2006}. However, as will be shown, the uncertainty on the input (\emph{i.e.} the measured volumetric errors $\boldsymbol{\delta \tau}_1 \cdots \boldsymbol{\delta \tau}_n$) cannot be modelled by a simple distribution.

Instead of choosing an arbitrary number of trials, considered sufficient, the simulation is \emph{adaptive}, as described in \cite{gum_mc}. The main purpose of implementing an adaptive method is to provide \emph{control variables} (\emph{i.e.} expected values $\overline{\boldsymbol{y}}$, standard uncertainties $u(\boldsymbol{y})$ and lower and higher coverage interval endpoints $\boldsymbol{y}_{low}$ and $\boldsymbol{y}_{high}$) with an expected numerical tolerance.

Briefly, the adaptive method consists in performing the Monte Carlo simulation divided into sequences of $M$ trials until the maximum of each standard deviations sets associated with the average of the estimate for each sequence $s_{\overline{y}}$, the standard uncertainty $s_{u(y)}$, and lower and higher coverage interval endpoints $s_{y_{low}}$ and $s_{y_{high}}$ are lower than the chosen numerical tolerance (see \cite{gum_mc} and Fig.~\ref{fig:structure_mc}). The algorithm described in \cite{gum_mc} has been implemented in MATLAB language in house.

In this case, $M$ has been set to $10^4$ for a coverage probability of 95\%. The numerical tolerance $\delta$ has been set to $0.5\cdot 10^{-4}\,mm$.

The model of uncertainty includes three sources: the drift of the closed kinematic chain during the measurement, the uncertainty on the sensors output values and the uncertainty due to the projection from the sensing head frame to the tool frame. Those uncertainty sources are part of the measurement and identification process.

Potential uncertainty due to the NC-unit action is removed using the exact stop option in the NC program. The influence of joint motion errors \cite{knap:2009} was not considered.

\subsection{Sensors}
Standard type A uncertainty on the sensors output has been evaluated in the useful measurement range, according to \cite{gum}. It is used to produce random values within a normal distribution in the Monte Carlo analysis for each point and sensor. The standard measurement uncertainties are given in Table~3.

\begin{table}
\center
\caption{Standard measurement uncertainty of the sensors.}
\label{tab:sigma_capteurs}
\vspace{4pt}
\small{
\begin{tabular}{p{6cm}c}
\emph{Sensor -- Channel}  &   \emph{Value}  \\
\hline
Sensor 4 -- Channel 1 &    $0.28 \, \mu m$  \\
Sensor 5 -- Channel 2 &    $0.28 \, \mu m$  \\
Sensor 6 -- Channel 3 &    $0.40 \, \mu m$  \\
\hline 
\end{tabular} 
}
\end{table}

\subsection{Transformation uncertainties}

\begin{figure}[tbh]
\center
\includegraphics[scale=1]{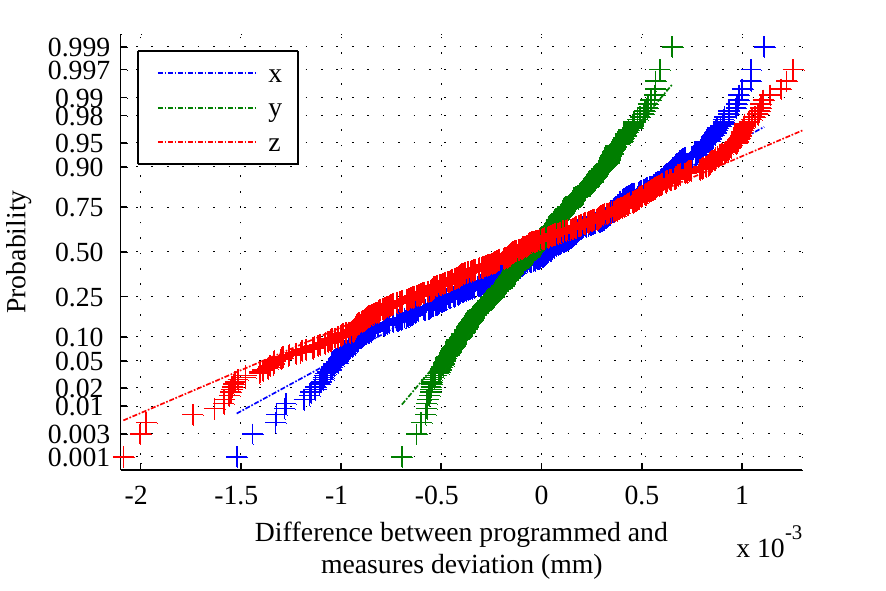}
\caption{Distributions of the difference in $x$, $y$ and $z$ directions between programmed and measured volumetric errors for 500 points in a normal probability plot -- Standard deviations $u_x=0.56 \, \mu m$, $u_y=0.27 \, \mu m$ and $u_z=0.69 \, \mu m$.}
\label{fig:dispersion_calibration}
\end{figure}

As described previously, the measurement result of each sensor is used to provide the measured volumetric error in the machine reference frame, using the transform matrix $\mathcal{M}_{s \rightarrow t}$. Standard uncertainties on the terms of this matrix induce additional uncertainties on the estimated parameters.
To evaluate this effect, the trajectory used for calibration was run four times to provide a total of 500 points where the small programmed X-, Y-, Z-axes motions can be compared to the measured ones. Distributions of the differences between programmed and measured errors are given in Fig.~\ref{fig:dispersion_calibration}.

The distributions observed correspond with normal distributions so the influence of the transformation from the sensors frame to the machine frame has been modelled by a normal distribution reflecting the evaluated standard deviations observed.

This model can be seen as slightly pessimistic, since this uncertainty model is based on data that are themselves suffering sensors uncertainties and thermal variations. The thermal variation effect can be neglected, considering the 4 tests have been performed within 3 minutes. No further hypothesis were made to avoid including the contribution of sensors uncertainties here. The modelled effect of the frame transformation is known to be over evaluated, leading to higher total uncertainties.

\subsection{Drift of the closed kinematic chain}

\begin{figure}[t]
\center
\includegraphics[scale=1]{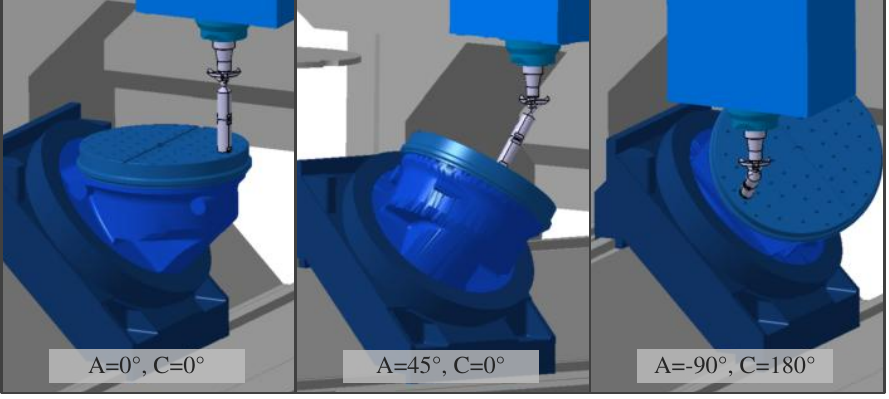}
\caption{CAD view of the 3 poses for the evaluation of the dimensional variation of the measurement chain.}
\label{fig:poses_statiques}
\end{figure}

\begin{figure}[t]
\center
\includegraphics[scale=1]{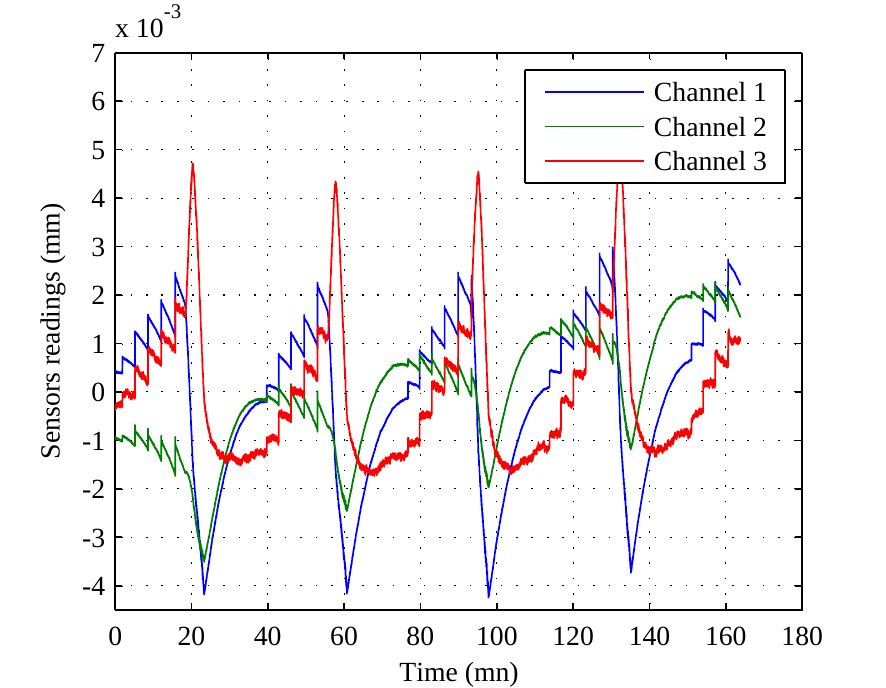}
\caption{Example of thermal variations measured during a static test.}
\label{fig:allures_cycles}
\end{figure}

The thermal expansion of the spindle during machining is a well known problem among the machining community \cite{schm:2008}. The dimensional variations -- or drift -- of the closed kinematic chain has been considered in the uncertainty model.

An example of the drift cycles is shown in Fig.~\ref{fig:allures_cycles}. The spindle temperature is regulated by a cooling device which generates thermal cycles. During those cycles, the spindle suffers small movements of approximately $7.5\, \mu m$ peak to valley in the $z$ direction and $4.5\, \mu m$ peak to valley in the $x$ and $y$ directions. Such a high magnitude of the drift may result from the use of the machine not under normal condition met during machining.

Similar were observed for three different poses, depicted in Fig.~\ref{fig:poses_statiques}. For those three poses, similar period, magnitude and pattern have been observed between channels. It suggests that the observed effect is not a variation of the mesurand (\emph{i.e.} the 8 link errors) but a variation in the measurement chain, as the observed volumetric variations do not vary with the pose, influenced itself by the link errors. However, if the machine link errors -- the mesurand -- are affected, it is still a contribution to uncertainty since it means that the desired conditions of the mesurand are not met. Such conditions are difficult to define for a machine tool since its metrology is not conducted during normal \emph{machining conditions}.

The drift is included in the uncertainty model. Two different methods are compared: on one hand the drift is treated statistically and on the other hand, its cyclic character is preserved in the model.

\subsubsection{Statistical method}

The principle given for linear positioning measurement in ISO/TR 230-9 \cite{iso230-9} is applied. It takes into account the magnitude of the drift -- called $E_{VE}$ -- to calculate the uncertainty due to environmental variation $u_{EVE}$ as described by eq.\eqref{eq:u_eve}.

\begin{equation}
u_{EVE}=\frac{E_{VE}}{2\sqrt{3}}
\label{eq:u_eve}
\end{equation}

In this case, a value of $u_{EVE}$ is evaluated for each sensor to produce random values within a normal distribution subtracted for each point in the Monte Carlo analysis. The values of $E_{VE}$ summarised in Table~4 are evaluated according to the measurements shown Fig.~\ref{fig:allures_cycles}. In the following, this is called the statistical method.\\

\begin{table}
\center
\caption{Magnitude $E_{VE}$ and uncertainty $u_{EVE}$ of the drift.}
\label{tab:eve}
\vspace{4pt}
\small{
\begin{tabular}{p{3cm}cc}
\emph{Sensor -- Channel}  & $E_{VE}$	&    $u_{EVE}$  \\
\hline
Sensor 4 -- Channel 1 &    $6.95 \, \mu m$ &    $2.00 \, \mu m$  \\
Sensor 5 -- Channel 2 &    $3.42 \, \mu m$ &    $0.99 \, \mu m$  \\
Sensor 6 -- Channel 3 &    $6.63 \, \mu m$ &    $1.91 \, \mu m$  \\
\hline 
\end{tabular} 
}
\end{table}

\begin{figure*}[htb]
\center
\includegraphics[scale=1]{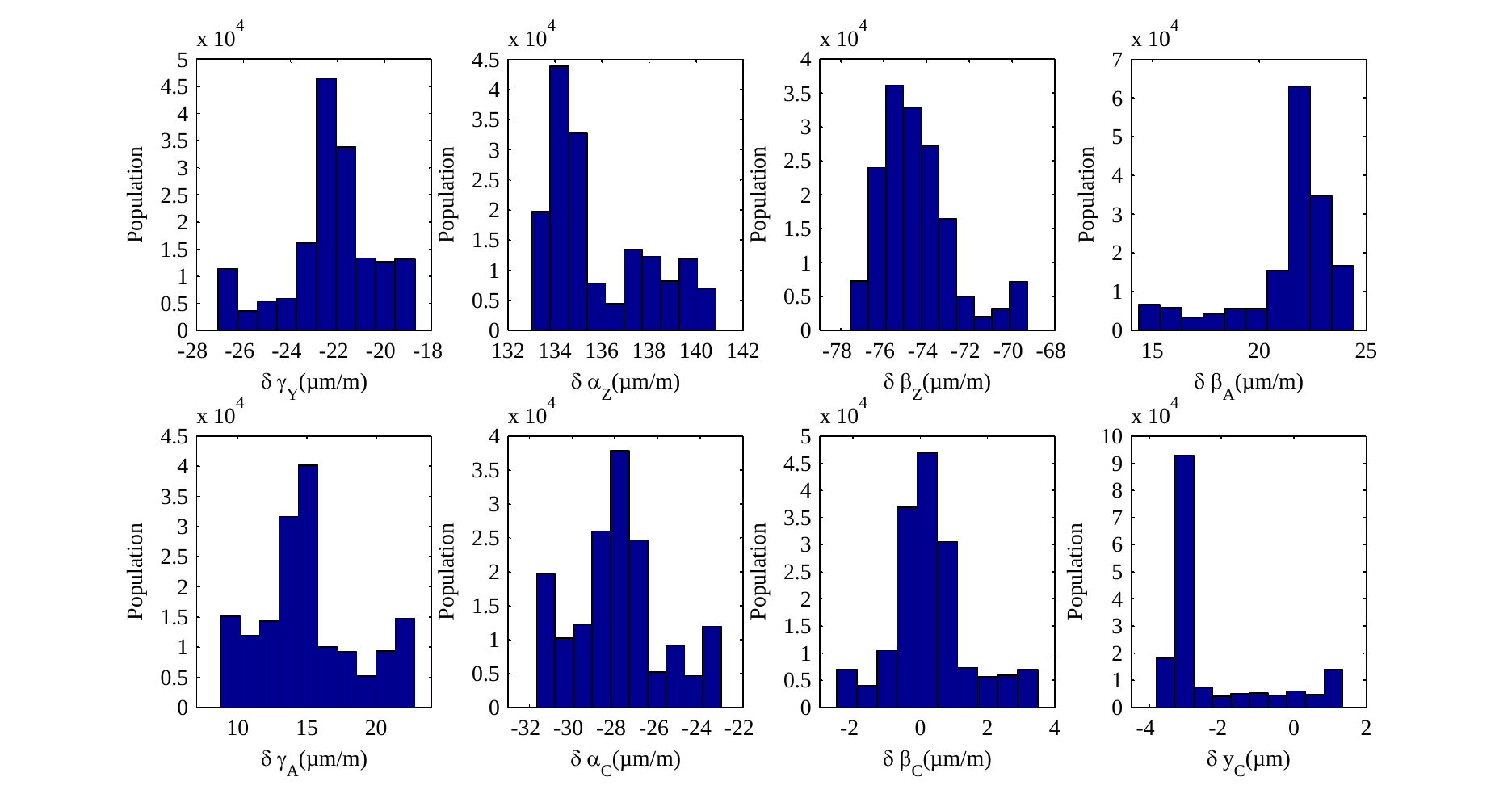}
\caption{Simulated distributions of the identified link errors with a 95\% coverage interval (cyclic method).}
\label{fig:output_mc}
\end{figure*}

\subsubsection{Cyclic method}

In the cyclic method, the uncertainty due to environmental drift is evaluated considering the cyclic character of the variation. A typical variation period of these cycles (Fig.~\ref{fig:allures_cycles}) has been chosen so that the drift is modelled as a periodical function of time $\boldsymbol{D}_{EVE}$ from $t=0$ to $+\infty$.

The beginning of the first trial of each sequence is chosen randomly in the period which gives the starting time $t_0$. The duration of a simulated trajectory is known as $t_m$, and the time interval between the volumetric errors measurement at two following point is known as $t_i$. Then, the measurement time $t_{nk}$ of the point number $k$ ($k$ varies from 1 to 807) in the trial number $n$ ($n$ varies from 1 to $M$) is given by eq.\eqref{eq:measurement_time}.

\begin{equation}
t_{nk}= t_0 + (n-1) \cdot t_m + k \cdot t_i
\label{eq:measurement_time}
\end{equation}

\noindent
where:
\begin{equation}
t_{m}= 807 \cdot t_i
\label{eq:t_m}
\end{equation}

The simulated drift at this point is given by $\boldsymbol{D}_{EVE}(t_{nk})$. The calculated variation is subtracted from the simulated measured value. In the following, this is called the cyclic method.

\section{Results}

\begin{figure*}[p]
\center
\includegraphics[scale=1]{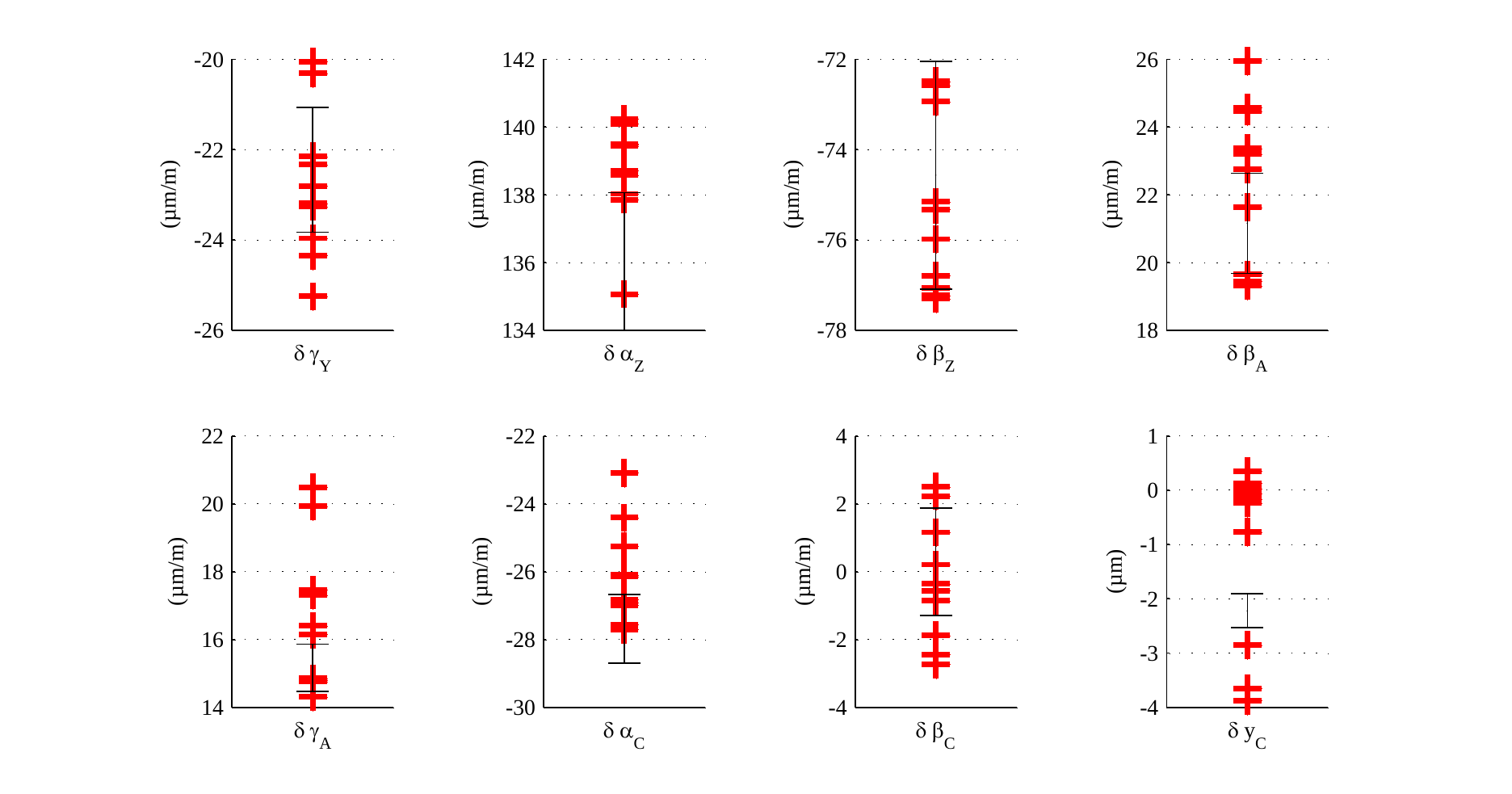}
\caption{Comparison between simulated 95\% coverage interval obtained with the statistical method \emph{(black or dark)} and experimentally identified link errors \emph{(red or grey)}.}
\label{fig:validation_mc_iso}
\end{figure*}

\begin{figure*}[p]
\center
\includegraphics[scale=1]{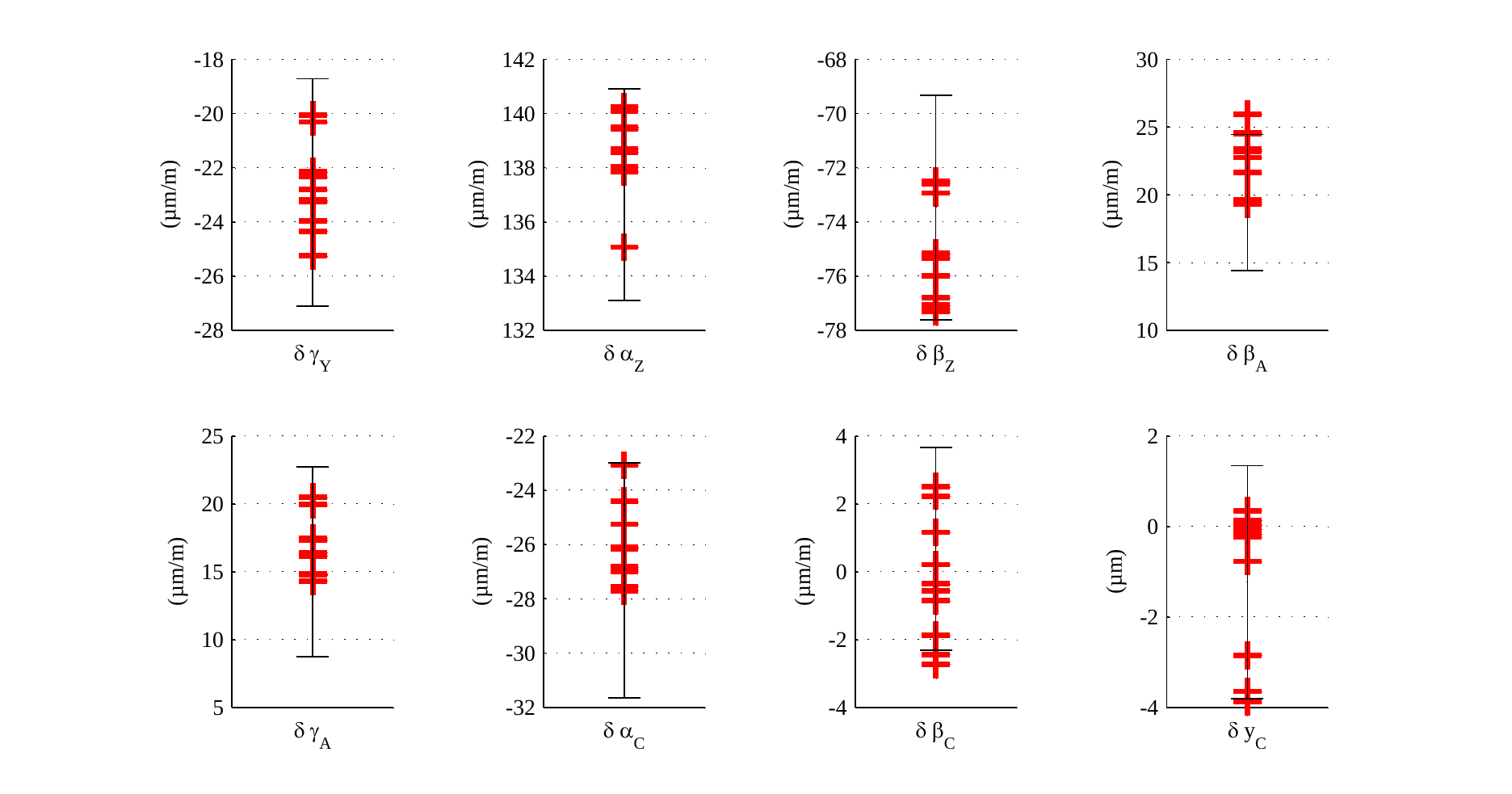}
\caption{Comparison between simulated 95\% coverage interval obtained with the cyclic method \emph{(black or dark)} and experimentally identified link errors \emph{(red or grey)}.}
\label{fig:validation_mc}
\end{figure*}

\subsection{Simulation results}

The reliability of the multi-output Monte Carlo simulation method was evaluated by confronting simulation results to experimental values: 15 identification trajectories were executed, leading to 15 error sets $\boldsymbol{\delta \chi}$. The first 5 sets were removed from the analysis to avoid including thermal variation of the joints, leading to potential variations of the mesurand. Then, eq.\eqref{eq:solution} provides an array of identified link and setup errors for each of the 10 tests considered as relevant. The execution of those 10 tests took approximately 3 hours, so several drift cycles occurred over this period (while the machine joints, assumed to be unaffected by the thermal drift, mainly attributed to the spindle, remains in a steady state after the 5 previous test warm up period).

The mean value of the 10 $\boldsymbol{\delta \chi}$ sets is used as an input for the Monte Carlo simulation. Other inputs are either the $u_{EVE}$ value for each sensor or the periodical function of the time $\boldsymbol{D}_{EVE}$, standard uncertainties on the sensors output and standard uncertainties of the sensor to machine frame transformation. Outputs of the simulation are the expected distribution of the identified link errors (Fig.~\ref{fig:output_mc}) and lower and higher coverage interval endpoints (Fig.~\ref{fig:validation_mc_iso} and Fig.~\ref{fig:validation_mc}). The 95\% coverage intervals given by the Monte Carlo simulation are compared to the 10 relevant tests results in Fig.~\ref{fig:validation_mc_iso} for the statistical method and in Fig.~\ref{fig:validation_mc} for the cyclic method. The mean values of Fig.~\ref{fig:validation_mc_iso} and Fig.~\ref{fig:validation_mc} do not exactly match the results presented in the Table~1 because identification results of the Table~1 were obtained during the winter and the the experiments leading to Fig.~\ref{fig:validation_mc_iso} and Fig.~\ref{fig:validation_mc} were performed during the summer in a laboratory where the global climate condition can fluctuate a lot. This statement shows that the machine geometry is also subject to larger  alterations on a long period, but this phenomenon is not discussed further in this paper.

The statistical method leads to uncertainty intervals smaller than observed variations of the identified errors, whereas the simulation with the cyclic method provides a more realistic size for the 95\% coverage interval for each link error. This illustrates the importance of taking into account the cyclic character of the drift in the model. From this point onward, only the cyclic method is kept in the analysis.

As shown in Fig.~\ref{fig:output_mc}, the measurement chain variation cycles lead to non-normal distributions\footnote{Simulation run without including the drift led to normal distributions.}, even if expected values correspond to identified values with the $\boldsymbol{\delta \chi}$ input. 

Sizes of the 95\% coverage intervals for the cyclic method are given for each link error in Table~5.

\subsection{Numerical stability of the cyclic method}

\begin{figure}
\center
\includegraphics{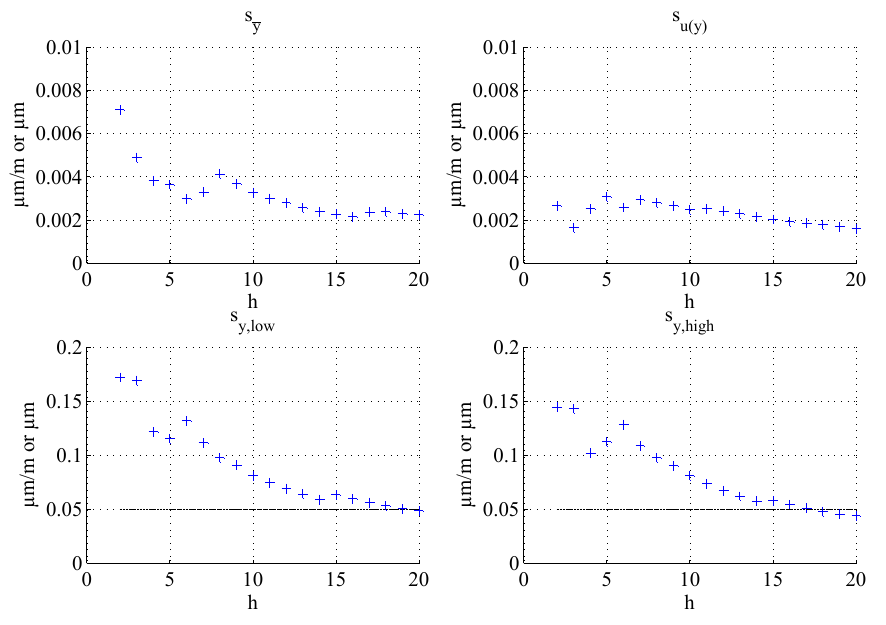}
\caption{Evolution and convergence of the control variables for a typical simulation -- $h$ is the Monte Carlo sequence number; the numerical tolerance is  $\delta = 0.05 \, \mu m \, or \, \mu m/m$.}
\label{fig:convergence_monte_carlo}
\end{figure}

Ten simulations were executed with the cyclic method to characterise the numerical stability. All ten simulations converged to the same outputs in terms of expected values, standard uncertainties and lower and higher endpoints of the 95\% coverage interval, with a $\pm \delta$ tolerance previously defined as $10^{-4}\,mm$. From this point of view, the method can be considered numerically stable.

A typical convergence of the control variables is shown in Fig.~\ref{fig:convergence_monte_carlo}. The number of sequences performed to reach the convergence criteria for all the control variables is $h=20$ in this case. The lower and higher endpoints of the 95\% coverage interval proved to require more sequences to reach convergence. For the ten simulations, the number of sequences before convergence varied from 17 to 24.

\subsection{Contribution of each uncertainty source}

\begin{table}[tbh]
\center
\caption{Size of the 95\% coverage interval for each uncertainty sources separately, compared to total 95\% coverage interval.}
\vspace{4pt}
\label{tab:contributions_mc}
\small{\begin{tabular}{p{1.55cm}ccccc}
\emph{Link error} & ${\Delta_{EVE}}^\text{a}$ & $\Delta_{sensors}$ & $\Delta_{trans}$ & $\sqrt{\sum \Delta^2}$ & ${\Delta_{95\%}}^\text{a}$ \\
\hline
\hline $\delta \gamma_Y \, (\mu m/m)$ &    8.1 	&  	0.49 	&  0.68 	& 8.1 	& 8.4  \\
\hline $\delta \alpha_Z \, (\mu m/m)$ &    7.4 	&  	0.65 	&  0.85 	& 7.5 	& 7.8  \\
\hline $\delta \beta_Z \, (\mu m/m)$  &    7.9 	&  	0.88 	&  1.41 	& 8.1 	& 8.2  \\
\hline $\delta \beta_A \, (\mu m/m)$  &    9.8 	&  	0.48 	&  0.87 	& 9.9 	& 10.0 \\
\hline $\delta \gamma_A \, (\mu m/m)$ &   13.9 	&  	0.29 	&  0.64 	& 13.9 	& 14.0 \\
\hline $\delta \alpha_C \, (\mu m/m)$ &    8.4 	&  	0.38 	&  0.75 	& 8.4 	& 8.6  \\
\hline $\delta \beta_C \, (\mu m/m)$  &    5.9 	&  	0.48 	&  0.81 	& 6.0 	& 6.0  \\
\hline $\delta y_C \, (\mu m)$  	  &    5.1 	&  	0.12 	&  0.21 	& 5.1 	& 5.2  \\
\hline
\end{tabular}}
\flushleft $^\text{a}$\emph{For the cyclic method.}
\end{table}

Three other simulations were performed this time, including only one source of uncertainty at a time. This allows to compare the contribution of each source of uncertainty. Table~5 shows the total contribution of all uncertainty sources $\Delta_{95\%}$ (for the cyclic method) compared to $\Delta_{EVE}$ for the drift variations only (cyclic method), $\Delta_{sensors}$ for the sensors output uncertainty only and $\Delta_{trans}$ for the frame transformation uncertainties only. Those simulated values show that the dimensional variations are responsible for most of the total uncertainties.

It is noticeable that for each error parameter, the quadratic sum of the 3 contributions $\sqrt{\sum \Delta^2}$ is lower than the predicted $\Delta_{95\%}$ probably a consequence of the cyclic character of the drift included in the model.

\section{Conclusion}

The paper presents a method to evaluate contributions to the uncertainties for identified link errors of a five axis machine tool. Sources of uncertainty in the identification procedure have been inventoried, considering certain hypothesis. Those standard uncertainties have been propagated with a multi-output adaptive Monte Carlo approach, using either a statistical model or a cyclic model for the drift.

The main source of uncertainty in the procedure, for the tested machine in the prevailing experimental conditions, was the drift of the closed kinematic chain. The cyclic nature of the drift proved to have a significant impact on the simulation results.

The numerical stability of the implemented method was evaluated, and the obtained uncertainties compared to results from ten experimental identification trials.\\

The method was also used to evaluate the contribution of each of the three uncertainty source in order to pinpoint the dominant source, in order to improve the link errors identification method by decreasing or removing its impact. The effect of environmental variation errors, generally attributed to thermal variations of the spindle, proved to be predominant under the conditions of the tests. The thermal stability control during the identification must now be improved to decrease the total uncertainty of the procedure.

\section*{Acknowledgements}

This work was partially funded with a Discovery Grant from the National Science and Engineering Research Council of Canada and conducted on equipment purchased with a grant from the Canadian Foundation for Innovation.

\end{document}